\documentclass[10pt,twosides,a4paper]{book}
\usepackage{bbm}
\usepackage{bm}
\usepackage{physics} 
\usepackage{amssymb}
\usepackage[pdftex]{graphicx}
\usepackage{latexsym}
\usepackage[small]{caption2}
\usepackage{amsmath}
\usepackage[english]{babel}
\usepackage{babel}
\usepackage{amsmath,amsthm}
\usepackage{rotating}
\usepackage{multirow}
\usepackage{graphicx}
\usepackage{tocloft}
\usepackage{bbm}
\usepackage{dsfont}
\usepackage{amsfonts}
\usepackage{accents}
\usepackage{pdfpages}
\usepackage[thinc]{esdiff}
\usepackage{hyperref}
\usepackage[numbers,sort&compress,square]{natbib}
\usepackage{natbib}
\usepackage{pdfsync}
\usepackage{fancyhdr}
\usepackage{scalerel,stackengine}
\usepackage{fancyhdr}
\usepackage{booktabs}
\renewcommand{\cftchappresnum}{CHAPTER }
\newlength{\mylen}
\fancypagestyle{plain}{} 

\input{tcilatex} 

\begin{document}

\chapter[Isa\'ias Vallejo-Fabila and Lea F. Santos \\
{\em Timescales for thermalization, quantum chaos, \\ and self-averaging}]{Timescales for thermalization, quantum chaos, and self-averaging}
\label{chapter8}

\markboth{Timescales for thermalization, quantum chaos, and self-averaging}{Isa\'ias Vallejo-Fabila and Lea F. Santos}

\noindent{\large \textbf{Isa\'ias Vallejo-Fabila$^{a}$ and Lea F. Santos$^{b}$}}

\vspace{3mm}

\noindent{Department of Physics, University of Connecticut, Storrs, Connecticut 06269, USA}

\noindent{\texttt{$^a$isaias.vallejo@uconn.edu; $^{b}$lea.santos@uconn.edu}}


\section{Introduction}

This chapter discusses the conditions and timescales under which isolated many-body quantum systems, initially far from equilibrium, ultimately reach thermal equilibrium. We also examine quantities that, during the relaxation process, exhibit dynamical manifestations of spectral correlations as in random matrix theory and investigate how these manifestations affect their equilibration times. We refer to systems presenting these spectral correlations as chaotic quantum systems, although the correct term to be employed, whether chaotic or ergodic quantum systems, is debatable and both have limitations.

The chapter is organized as follows:
\begin{itemize}
\item \emph{Thermalization in isolated many-body quantum systems:} We begin with an introduction to thermalization in isolated many-body quantum systems within the framework of the eigenstate thermalization hypothesis (ETH) \cite{vall.Alessio2016}, using full random matrices as a reference and comparing them with physical systems~\cite{vall.Borgonovi2016}.

\item \emph{Thermalization timescales and correlation hole:} Next, we discuss the timescales that chaotic quantum systems take to thermalize, investigating the dependence on system size, model, and quantities. The intuition that larger systems equilibrate faster is confirmed for the participation entropy evolving under full random matrices~\cite{vall.TorresEntropy2016}. However, for chaotic spin-1/2 chains with nearest-neighbor couplings and onsite disorder, the participation entropy exhibits polynomial equilibration timescales~\cite{vall.Lezama2021}. Furthermore, quantities that reveal dynamical manifestations of quantum chaos~\cite{vall.Torres2017Ann,vall.Torres2017Philo,vall.Torres2018,vall.Santos2020}, known as correlation hole~\cite{vall.Leviandier1986,vall.Guhr1990,vall.Gorin2002,vall.Wilkie1991,vall.Alhassid1992}, require an exponentially long time to equilibrate~\cite{vall.Torres2017Ann,vall.Torres2017Philo,vall.Torres2018,vall.Santos2020,vall.Schiulaz2019,vall.Lezama2021}, independently of whether we consider random matrices or physical models. This delay arises because the dynamics must resolve the discreteness of the spectrum to detect spectral correlations. The timescale for such quantities to equilibrate is the Heisenberg time -- the inverse of the mean level spacing -- which represents the longest timescale in quantum dynamics. Despite these long timescales, we explain that dynamical manifestations of spectral correlations could be observed in current experiments using cold atoms \cite{vall.Schreiber2015} and ion traps \cite{vall.Richerme2014}, or with commercially available quantum computers if small many-body quantum systems are considered~\cite{vall.Das2025}.

\item \emph{Self-averaging in closed and open quantum systems:} Quantities that exhibit dynamical manifestations of spectral correlations are non-self-averaging~\cite{vall.Schiulaz2020}, which means that their time evolution displays large fluctuations that persist with increasing system size. Consequently, large sample sizes are required to average out fluctuations and reveal spectral correlations. We demonstrate how coupling the system to an energy dephasing environment can reduce fluctuations, ensuring self-averaging and the visibility of the correlation hole with few disorder realizations.

\end{itemize}

\section{Thermalization}

To explain thermalization in isolated many-body quantum systems, we begin by examining the properties of full random matrices, where thermalization is straightforward. Full random matrices  are filled with random numbers from a Gaussian distribution. We consider random matrices from the Gaussian orthogonal ensemble (GOE), which are real and symmetric, aligning with the Hamiltonians of the isolated physical systems that we study. The elements of GOE random matrices satisfy  $\langle H_{ij} \rangle =0$ and

\begin{equation}
\left\langle H^{2}_{ij} \right\rangle =\left \{
\begin{array}{r}
1\hspace{1.7cm} i=j , \\
1/2\hspace{1.5cm} i\ne j .
\end{array}
\right.
\end{equation}
The eigenstates of these matrices are random vectors constrained by normalization. This means that the components $C_{\alpha}^{k}$ of any eigenstate $|\alpha \rangle$ are Gaussian distributed random numbers satisfying $\sum_{k=1}^D |C_{\alpha}^{k}|^2$, where $D$ is the matrix dimension. Therefore,  all eigenstates are statistically equivalent.

The eigenvalues of GOE matrices are highly correlated. This applies not only to adjacent levels, leading to the Wigner-Dyson distribution, $P(s) = (\pi s/2)\exp(-\pi s^{2}/4)$, of the spacings $s$ between neighboring levels, but also to distant energy levels, which can be verified with the level number variance, rigidity, or spectral form factor~\cite{vall.Guhr1998,vall.MehtaBook}. The spectrum of full random matrices is rigid, and degeneracies are avoided.

Let us analyse the evolution of a given observable $O$,
\begin{equation}
\langle O (t) \rangle = \sum_{\alpha \neq \beta} C_{\beta }^{k_0*} C_{\alpha}^{k_0} e^{i(E_{\beta} - E_{\alpha})t} \langle \beta |O| \alpha \rangle + \sum_{\alpha} |C_{\alpha}^{k_0}|^2 \langle \alpha |O| \alpha \rangle,
\end{equation}
under GOE random matrices. The picture is that of quench dynamics, where the initial state $|\Psi(0)\rangle$ corresponds to a state $|k_0 \rangle $ of the diagonal part of the matrix, $H_0$, which then evolves under the full Hamiltonian $H=H_0 + V$, with $V$ representing the off-diagonal elements. Due to spectral rigidity and the large matrix sizes, the first term in the equation above averages out over long times. We could then proceed to analyze whether the infinite-time average (diagonal ensemble), $\overline{O} = \sum_{\alpha} |C_{\alpha}^{k_0}|^2 \langle \alpha |O| \alpha \rangle$, coincides with a thermodynamic average, as required for thermalization. But before proceeding, it is important to take a closer look at the first term of the equation above, as it can anticipate what to expect for the infinite-time average. Since the eigenstates of $H$ are random vectors, any initial state projected in this basis is also a random vector, meaning $C_{\alpha}^{k_0}$ are random numbers. Moreover, because the eigenstates are random vectors, the off-diagonal elements $\langle \beta |O| \alpha \rangle$ follow a Gaussian distribution~\cite{vall.Beugeling2015,vall.LeBlond2019,vall.Santos2020}. This analysis of eigenvalues, initial state components, and matrix elements $\langle \beta |O| \alpha \rangle$ provides insights into whether thermalization will occur, even before explicitly evaluating $\overline{O}$. It also allows to determine whether at long times, $\langle O (t) \rangle$ approximates $\overline{O}$ for most times, with fluctuations decreasing exponentially as the system size increases~\cite{vall.Zangara2013,vall.Lezama2023}.

After the study of the first term, we complete the analysis by assessing how closely the infinite-time average $\overline{O}$ aligns with the microcanonical average, $ O_{micro}$, that is,
\begin{equation}
\sum_{\alpha} |C_{\alpha}^{k_0}|^2 \langle \alpha |O| \alpha \rangle \approx  \frac{1}{{\cal N}} \hspace{-0.6 cm } \sum_{\substack{\alpha \\ |E_0 - E_{\alpha} |<\Delta E}}
\hspace{-0.5 cm } \langle \alpha | O | \alpha \rangle,
\end{equation}
and, crucially, whether this agreement improves as the system size grows. In the equation above, ${\cal N}$ is the number of states in the energy window $\Delta E$  and $E_0$ is the energy of the initial state. The  infinite-time average  $\overline{O}$ depends on the initial state through the components $|C_{\alpha}^{k_0}|^2$, but, as said above, its components are random numbers. Furthermore, since all eigenstates of GOE random matrices are statistically similar, the eigenstate expectation value, $\langle \alpha |O| \alpha \rangle$, computed with any of them gives equivalent results, apart from small fluctuations that decrease with system size. Therefore, the expectation value obtained with any single eigenstate should be close to the average, which is the basic idea of the eigenstate thermalization hypothesis (ETH). For these reasons, the left and right sides of the equation above are very similar and become even closer as the system size increases.

However, analysing thermalization in physical systems introduces additional subtleties. Unlike full random matrices, the Hamiltonian matrices of real systems are banded and sparse, and even if the system involves randomness, the elements are correlated. As a result, even in chaotic systems where level statistics resemble those of random matrices, their eigenstates are not completely random vectors. Nevertheless, away from the edges of the spectrum, the eigenstates approximate random vectors by filling the energy shell~\cite{vall.Santos2012PRE}. At the same time, scaling analyses of the participation ratio of the eigenstates $|\alpha \rangle$ away from the edges of the spectrum,
\begin{equation}
PR_{\alpha} = \frac{1}{\sum_{k=1}^{D} |C_{\alpha}^{k}|^4}
\end{equation}
show that it grows with the Hilbert space dimension~\cite{vall.Torres2017Ann,vall.PRRL2021}, indicating that the states are indeed delocalized. Therefore, thermalization is possible in chaotic systems, provided the initial state has energy away from spectral edges and the observables are local, so that they cannot detect the deviations from fully random eigenstates. [The behaviour of non-local observables in the context of thermalization has been discussed in~\cite{vall.Khaymovich2019}.]

\section{Timescales for thermalization}

\begin{figure}[!htb]
\begin{center}
\includegraphics[width=0.98\textwidth]{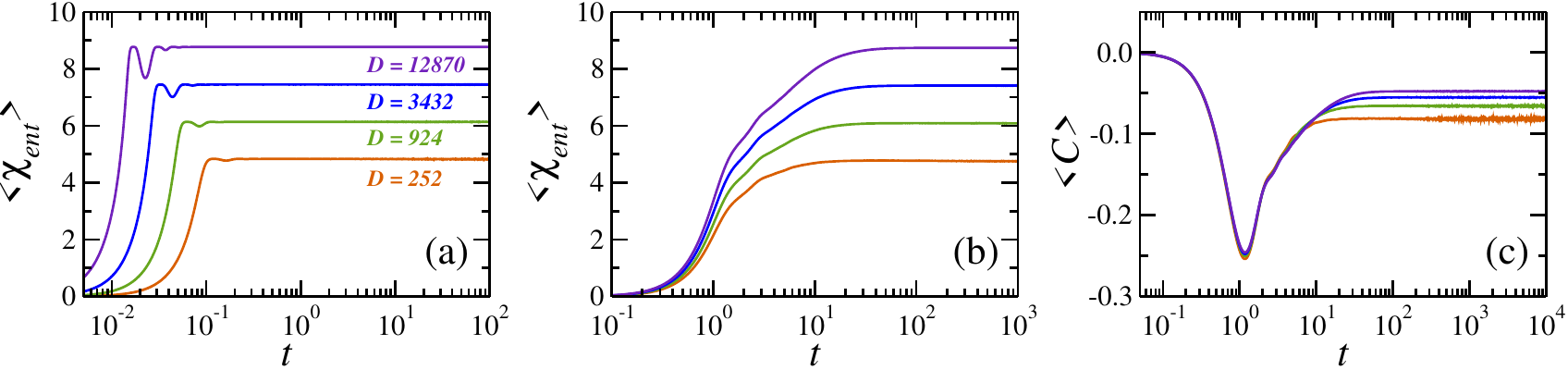} {}
\end{center}
\caption{Evolution of the participation entropy for (a) GOE random matrices of different sizes and (b) the disordered spin-1/2 chain in the chaotic regime ($h=0.75$). (c) Evolution of the spin-spin correlation function for the same model as in (b). The Hilbert space size for the spin model in (b)-(c) is the same as the matrix size in (a), which means $L=10$, $12$, $14$, and $16$, since we work in the subspace with ${\cal S}_z=0$. Averages are performed over disorder realizations and initial states. Initial states are close to the middle of the spectrum. }
\label{vall.fig01}
\end{figure}

We now investigate the timescales for thermalization in isolated chaotic many-body quantum systems.  To set the stage, we first examine the equilibration timescales for GOE random matrices. In Fig.~\ref{vall.fig01}(a), we present the evolution of the participation entropy~\cite{vall.Santos2012PRL,vall.Santos2012PRE},
\begin{equation}
\langle \chi_{\text{ent}} (t) \rangle = \left\langle - \ln \left( \sum_{k=1}^D |C_{k}^{k_0}(t)|^4 \right) \right\rangle
\end{equation}
up to saturation for different matrix sizes. Here, $\langle . \rangle$ indicates average over disorder realizations and initial states. The participation entropy measures the spreading of the initial $|\Psi(0)\rangle = |k_0\rangle$ over all basis vectors $|k\rangle$, that is, it quantifies the spreading of the initial state across the many-body Hilbert space. The figure shows that the entropy grows linearly (implying an exponential increase of the participation ratio~\cite{vall.Borgonovi2019PRER}) leading to rapid saturation of the dynamics. The equilibration time decreases as the matrix dimension increases, which aligns with our intuition that larger complex systems should equilibrate faster.

To compare with physical many-body systems, we consider the spin-1/2 Heisenberg model with nearest-neighbour couplings commonly studied in the context of many-body localization~\cite{vall.SantosEscobar2004,vall.Dukesz2009}
\begin{equation}
H = \sum_{k=1}^{L} h_{k} S_{k}^{z} + J \sum_{k=1}^{L}(S_{k}^{x}S_{k+1}^{x}+S_{k}^{y}S_{k+1}^{y}+S_{k}^{z}S_{k+1}^{z}),
\label{vall.Eq:Ham}
\end{equation}
where $h_{k}\in[-h,h]$, $h$ is the on-site disorder strength, $J$ is the interaction strength, and periodic boundary conditions are imposed. The model conserves total spin in the $z$-direction, ${\cal S}_z = \sum_n S_n^z$ and $[H,{\cal S}_z]=0$. We consider the subspace with ${\cal S}_z=0$, which means Hamiltonian matrices of size $D=L!/(L/2!)^2$.

In Fig.~\ref{vall.fig01}(b), we show the evolution of the participation entropy in the chaotic regime ($h=0.75$) for different system sizes. While the initial growth remains linear, similar to the GOE case, it slows down at later times, resulting in an equilibration time that increases with system size. Scaling analysis in Ref.~\cite{vall.Lezama2021} suggests that this time grows polynomially with the system size,  $t \propto L^{\gamma}$ with $\gamma>2$.

Since the participation entropy is a non-local quantity, we also analyse the spin-spin correlation function,
\begin{equation}
 \langle C(t) \rangle = \left\langle\frac{4}{L}\sum_{k=1}^{L}[\langle\Psi(t)|S_{k}^{z}S_{k+1}^{z}|\Psi(t)\rangle - \langle\Psi(t)|S_{k}^{z}|\Psi(t)\rangle \langle\Psi(t)|S_{k+1}^{z}|\Psi(t)\rangle]\right\rangle,
\end{equation}
which is experimentally studied in ion-trap systems~\cite{vall.Richerme2014}. Similar to the participation entropy, the equilibration time for $\langle C(t) \rangle$ also scales polynomially with $L$.

These results highlight that even in chaotic systems, equilibration times depend strongly on model-specific features. Identifying physical models that approximate the fast equilibration of random matrices is an interesting open question. Key aspects to explore include the range of interactions, system dimensionality, spin size, choice of observables, initial states, and symmetries.

A possible explanation for the slow thermalization observed in Fig.~\ref{vall.fig01}(b)-(c) is the presence of symmetries. Our system conserves total spin in the $z$-direction, and the observable $\langle C(t) \rangle$ involves spin operators in the same direction. Similarly, the participation entropy is computed in the eigenbasis of ${\cal S}_z$. Studies have shown that the relaxation of out-of-time-ordered correlators (OTOCs) can be slow even in chaotic systems when the OTOC operators overlap with the Hamiltonian~\cite{vall.Balachandran2023}. While slow relaxation does not necessarily imply that the equilibration time should grow with the system size, it hints at a possible connection. A definitive answer requires further numerical studies and potentially new theoretical insights.

Another recent development in thermalization studies that may shed light on slow dynamics is the concept of non-Abelian ETH~\cite{vall.Srednicki2023}. It refines the notion of microcanonical ensembles for systems with noncommuting conserved quantities. A key example is systems with SU(2) symmetry, where it has been suggested that equilibration times may be long. However, whether this leads to a system-size-dependent growth in equilibration time remains an open question.

\section{Correlation hole}

\begin{figure}[!htb]
\begin{center}
\includegraphics[width=0.98\textwidth]{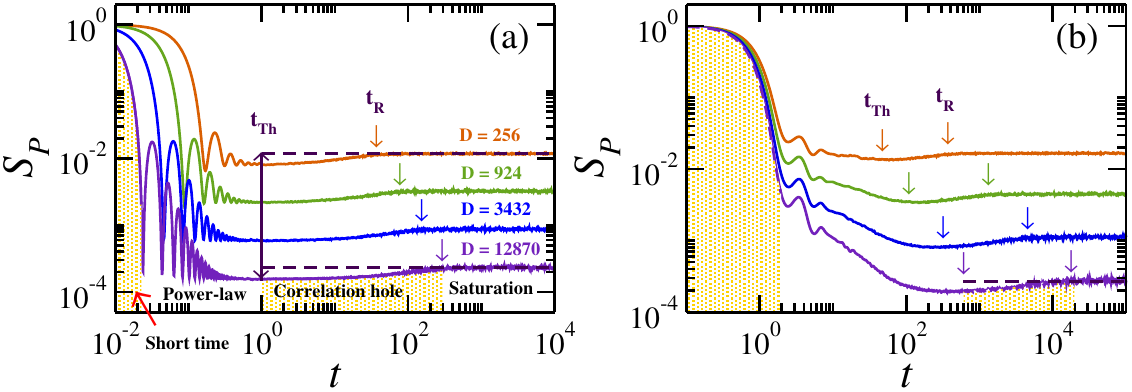} {}\\
\end{center}
\caption{Evolution of the survival probability for (a) GOE matrices of different sizes and (b) the disordered spin-1/2 chain in the chaotic regime ($h=0.5$) for initial states close to the middle of the spectrum. The Hilbert space size for the spin model in (b) is the same as the matrix size in (a), which means $L=10$, $12$, $14$, and $16$, since we work in the subspace with ${\cal S}_z=0$. In both panels $t_\text{R}$ indicates the equilibration time and $t_\text{Th}$ is the time for the minimum of the correlation hole, where the ramp starts. Averages are performed over disorder realizations and initial states.}
\label{vall.fig02}
\end{figure}

In addition to the model and its symmetries, the equilibration time also depends on the observables. To illustrate this, we once again consider the featureless case of full random matrices and compare the dynamics of the participation entropy in Fig.~\ref{vall.fig01}(a) with the evolution of the survival probability,
\begin{equation}
\langle S_p(t) \rangle = \left\langle | \langle \Psi(0)|\Psi(t) \rangle   |^2  \right\rangle = \left\langle \sum_{\alpha , \beta} |C_{\beta }^{k_0}|^2 |C_{\alpha}^{k_0}|^2 e^{i(E_{\beta} - E_{\alpha})t} \right\rangle
\end{equation}
shown in Fig.~\ref{vall.fig02}(a). Given that we are dealing with full random matrices, an analytical expression for the evolution of the survival probability can be derived~\cite{vall.Torres2018,vall.Schiulaz2019}
\begin{equation}
\langle S_p(t) \rangle = \frac{1-\overline{S_p}}{D-1} \left[ D\frac{\mathcal{J}_{1}^{2}(2\Gamma t)}{(\Gamma t)^{2}} -b_{2}\left(\frac{\Gamma t}{2D}\right) \right] + \overline{S_p}
\label{vall.Eq:SPanalytical}
\end{equation}
where $\mathcal{J}_{1}(t)$ is the Bessel function of first kind,
\begin{equation}
b_2(t) = [1-2t+t\ln(1+2t)]\Theta(1-t) + [t\ln[(2t+1)/(2t-1)]-1]\Theta(t-1)
\end{equation}
is the two-level form factor~\cite{vall.MehtaBook}, $\Theta(t)$ is the Heaviside step function, and
\begin{equation}
\overline{S_p} = \sum_{\alpha} |C_{\alpha}^{k_0}|^4 = IPR_0
\label{vall.Eq:IPR0}
\end{equation}
represents the saturation value. Here, $IPR_0$ denotes the inverse participation ratio of the initial state projected onto the energy eigenbasis. For GOE matrices, $IPR_0 \approx 3/D$. Figure~\ref{vall.fig02}(a) shows that after an initial power-law decay $\propto t^3$, which emerges from the first term in Eq.~(\ref{vall.Eq:SPanalytical}), the survival probability dips below the saturation value. This phenomenon, known as the correlation hole~\cite{vall.Leviandier1986,vall.Guhr1990,vall.Gorin2002,vall.Wilkie1991,vall.Alhassid1992,vall.DasG2023}, arises only in systems with correlated eigenvalues.  The correlation hole exhibits a ramp toward the saturation point, which has motivated the recent denomination ``dip-ramp-plateau structure''. The analytical expression in Eq.~(\ref{vall.Eq:SPanalytical}) allows us to determine the time for the beginning of the ramp, $t_\text{Th}$, which is constant for full random matrices, and the time for saturation, $t_\text{R}$, which is given by the inverse of the mean level spacing~\cite{vall.Schiulaz2019,vall.DasArXiv2025}. This equilibration time corresponds to the Heisenberg time, which scales with the size of the Hilbert space. Consequently, observables that exhibit dynamical manifestations of spectral correlations, what we refer to as dynamical signatures of quantum chaos, require an exponentially long time to reach equilibrium, even when evolving under full random matrices. This stands in stark contrast to the participation entropy dynamics shown in Fig.~\ref{vall.fig01}(a), where the saturation time decreases as the matrix size increases.

Before comparing the results in Fig.~\ref{vall.fig02}(a) to the survival probability of the chaotic physical model in Eq.~(\ref{vall.Eq:Ham}), we take a moment to discuss the relationship between survival probability and the spectral form factor,
\begin{equation}
\langle S_{FF}(t)  \rangle = \left\langle \frac{1}{D^2} \sum_{\alpha , \beta} e^{i(E_{\beta} - E_{\alpha})t} \right\rangle ,
\end{equation}
a quantity that has recently attracted significant attention. The spectral form factor is used to analyse spectral correlations in the time domain, but is not an actual dynamical quantity. If the initial state in the survival probability had equal components, $|C_{\alpha}^{k_0}|^2=1/D$, then $S_p(t)$ and $S_{FF}(t)$ would coincide. However, such an initial state is not physically realizable in conventional experimental setups. In laboratory conditions, such as those involving cold atoms and ion traps, initial states typically correspond to product states where each spin points either up or down in the $z$-direction, such as $|\uparrow \downarrow \ldots  \downarrow \uparrow \rangle$. These are the states, with energy in the middle of the spectrum, that we consider in our analysis of the survival probability in  Fig.~\ref{vall.fig02}(b).

As in the GOE case, the survival probability for the chaotic spin model in Fig.~\ref{vall.fig02}(b) requires Heisenberg time to saturate, meaning its equilibration time grows exponentially with system size $L$, that is,  $t_\text{R} \propto D/\Gamma$, where $\Gamma$ is the width of the initial state's energy distribution. However, unlike the GOE case, the time for the onset of the ramp in $\langle S_p(t) \rangle$ grows with system size~\cite{vall.Schiulaz2019}. Using the derivation of Eq.~(\ref{vall.Eq:SPanalytical}) as a reference, a semi-analytical expression for the entire evolution of $\langle S_p(t) \rangle$ under the chaotic spin model can be obtained (see Refs.~\cite{vall.Torres2018,vall.Schiulaz2019} for the analytical expression and \cite{vall.Tavora2016,vall.Tavora2017} for further discussions on the initial power-law decay of the survival probability). This expression shows that the time $t_\text{Th}$ for the beginning of the ramp scales as $t_\text{Th} \propto D^{2/3}/\Gamma$, as confirmed numerically.

The survival probability can be measured experimentally~\cite{vall.Das2025}, but as a nonlocal quantity, it is more challenging to measure compared to local observables.  Interestingly, numerical studies have shown that the spin autocorrelation function,
\begin{equation}
\langle I (t)  \rangle = \left\langle \frac{4}{L}\sum_{k=1}^{L}\langle\Psi(0)|S_{k}^{z}e^{iHt}S_{k}^{z}e^{-iHt}|\Psi(0)\rangle \right\rangle
\end{equation}
which is a local quantity, also detects the correlation hole~\cite{vall.Torres2018,vall.Schiulaz2019}. The difference with respect to the survival probability is that the correlation hole for $\langle I (t)  \rangle$ diminishes as the system size increases.

In nuclear physics, level statistics can be directly accessed through energy spectra. However, in experimental platforms such as cold atoms, ion traps, and quantum computers, access to spectral data is often limited. Instead, these platforms routinely simulate dynamical quantities, motivating the use of survival probability and spin autocorrelation to probe spectral correlations. The challenge, as seen in  Fig.~\ref{vall.fig02}(b), is that the correlation hole emerges only at long times and at very small values of $\langle S_p(t) \rangle$ and $\langle I (t)  \rangle$. To circumvent this, one can use small system sizes.

As shown in Ref.~\cite{vall.Das2025} even system sizes with only six sites exhibit the correlation hole. A relatively shallow quantum circuit was also proposed in Ref.~\cite{vall.Das2025} for the detection of the correlation hole using quantum computers. By running the circuit in a fake noisy IBM provider, the beginning of the ramp was observed. More recently, an experiment with superconducting quantum processors successfully measured the correlation hole~\cite{vall.Dong2025}.

\section{Self-averaging in open systems}

This final section addresses the issue of lack of self-averaging. An observable $O$ is self-averaging if its relative variance,
\begin{equation}
R_{O}(t) = \frac{\sigma_{O}^{2}(t)}{\langle O(t) \rangle^{2}},
\end{equation}
decreases as the system size increases. This allows for reducing the number of samples in numerical and experimental simulations as the system size increases. As early as the 1990s, numerical studies demonstrated that the spectral form factor is non-self-averaging~\cite{vall.Prange1997,vall.Braun2015}. More recently, it was shown analytically that both the survival probability and the spectral form factor are non-self-averaging~\cite{vall.Schiulaz2020} even for full random matrices. The same holds for the spin autocorrelation function for finite systems at timescales where the correlation hole emerges~\cite{vall.Schiulaz2020,vall.Schiulaz2020b}. This implies that regardless of system size, a large number of disorder realizations is required for the correlation hole to be observable. Without averaging, fluctuations obscure the dip-ramp-plateau structure.

In~\cite{vall.delCampo2021}, it was shown that the fluctuations in the spectral form factor decrease if the system is opened to a dephasing environment. For small dephasing strengths, even a single realization can reveal the dip-ramp-plateau structure. However, if dephasing is too strong, the correlation hole disappears, as it is a purely quantum feature linked to the discreteness of the spectrum.

\begin{figure}[!htb]
\begin{center}
\includegraphics[width=0.94\textwidth]{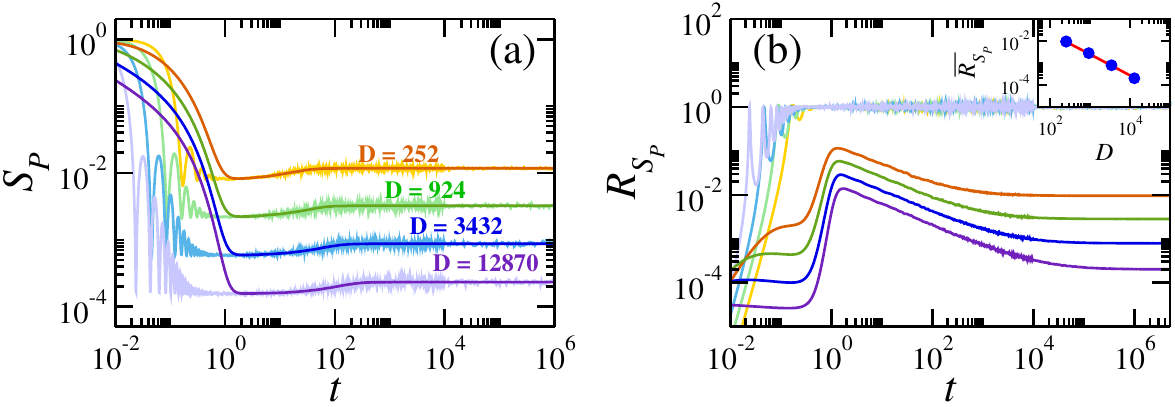} {}\\
\includegraphics[width=0.97\textwidth]{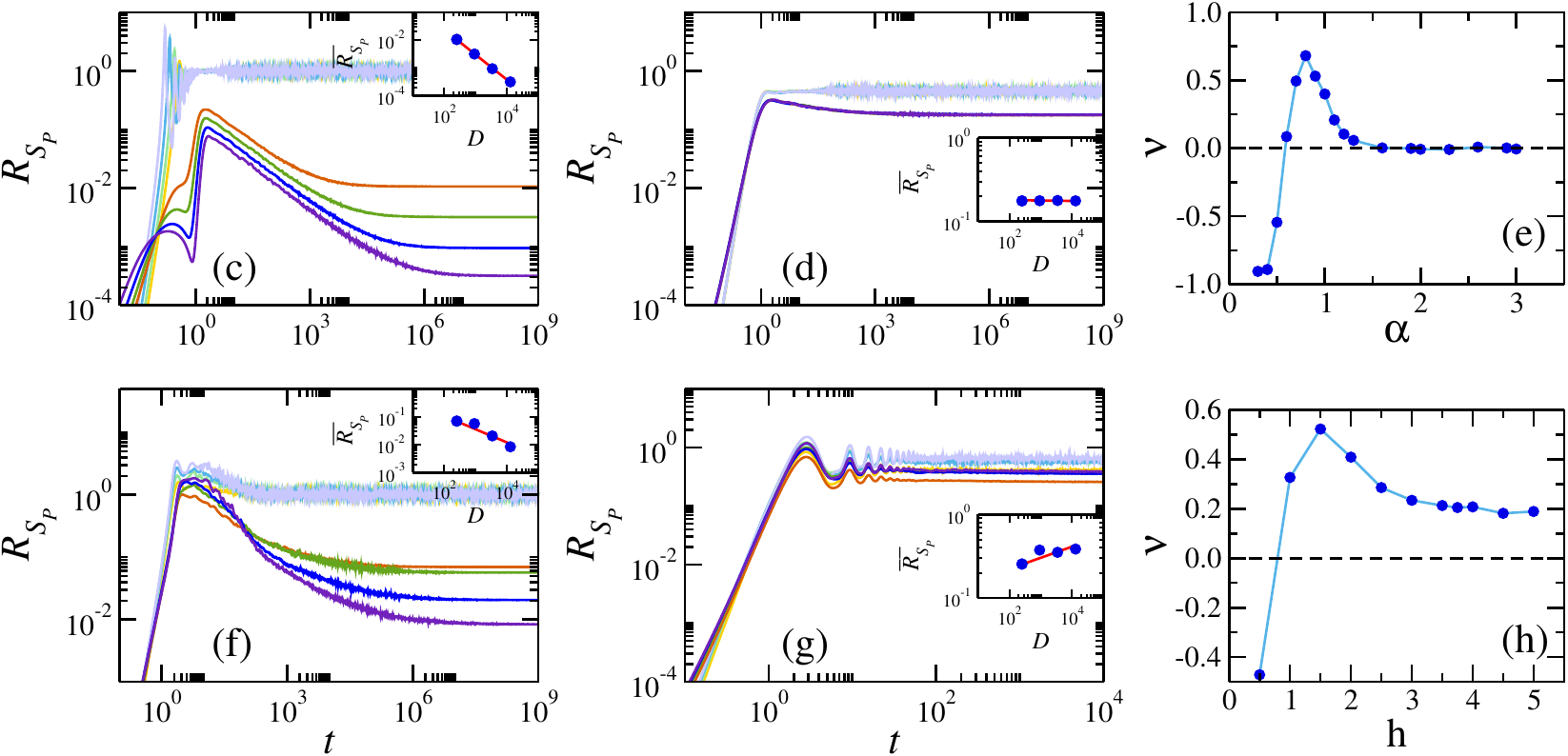} {}\\
\end{center}
\caption{Dynamics of the survival probability (a) and relative variance (b) for GOE matrices. Relative variance for power-law banded random matrices (PBRMs) (c) in the delocalized regime ($\alpha=0.3$) and (d) the localized regime ($\alpha=3.0$); (e) slope $\nu$ of the scaling analysis of the saturating value of the relative variance as a function of the parameter $\alpha$. Relative variance for a disordered spin model (f) in the delocalized regime ($h=0.5$) and (g) the localized regime ($h=5.0$); (h) slope $\nu$ of the scaling analysis of the saturating value of the relative variance as a function of the disorder strength. The Hilbert space size for the PBRM (c)-(e) and spin (f)-(h) models in is the same as the matrix size in (a)-(b), which for the spin model means $L=10$, $12$, $14$, and $16$, since we work in the subspace with ${\cal S}_z=0$. Insets in (b)-(d), (f)-(g) show the scaling analysis of the saturating value of the relative variance with the dimension $D$. Initial states are close to the middle of the spectrum. Averages are performed over disorder realizations and initial states. Light colours are for the isolated case while dark colours are for the open case  with $\kappa=0.05$.}
\label{vall.fig03}
\end{figure}

Inspired by the findings in~\cite{vall.delCampo2021}, the analysis of self-averaging for the survival probability was performed in~\cite{vall.Matsoukas2023} using  GOE random matrices and initial states corresponding to Gibbs states. Scaling analysis of the saturation values of $R_{S_{p}}(t)$ confirmed  that coupling to a dephasing environment not only reduces fluctuations, but also ensures self-averaging. The work draws an analogy between averages and dephasing due to an environment, where both lead to equivalent non-unitary dynamics. Essentially, opening the system mimics averaging, effectively smoothing the curves for the survival probability.

Despite these promising results, the question remained whether dephasing could also ensure self-averaging in physical models, motivating further studies in~\cite{vall.Vallejo2024}. The results are summarized in  Fig.~\ref{vall.fig03}, where the light lines correspond to isolated systems and dark lines to open systems. For a system coupled to a dephasing environment, the survival probability takes the form~\cite{vall.Matsoukas2023,vall.Vallejo2024,vall.Cornelius2022,vall.Matsoukas2024}
\begin{equation}
\langle S_p (t) \rangle = \left\langle \sum_{\alpha, \beta}|c_{\alpha}^{(0)}|^{2}|c_{\beta}^{(0)}|^{2}e^{-i(E_{\alpha}-E_{\beta})t-\kappa (E_{\alpha}-E_{\beta})^{2}t} \right\rangle,
\end{equation}
where $\kappa$ is the dephasing strength.

In Fig.~\ref{vall.fig03}(a), we compare numerical results for the survival probability evolving under GOE matrices for the isolated (light colours) and open (dark colours) system. Every curve is obtained after averages, but it is visible that the fluctuations are smaller in the open case. In Fig.~\ref{vall.fig03}(b), we show numerical results for $R_{S_{p}}(t)$ using GOE random matrices. For the isolated case, the relative variance for short times increases as $D$ increases and is independent of the matrix size at long times, confirming the lack of self-averaging at any timescale. Instead, by opening the system, the relative variance  decreases as $D$ increases throughout the dynamics. At equilibration, the relative variance is given by
\begin{equation}
\overline{R_{S_{p}}} = \frac{\sigma_{IPR_{0}}^{2}}{\langle IPR_{0} \rangle^{2}},
\end{equation}
where $IPR_0$ is the inverse participation ratio of the initial state in the energy eigenbasis [see Eq.~(\ref{vall.Eq:IPR0})]. Scaling analysis of $\overline{R_{S_{p}}} \propto D^{\nu}$ reveals that the slope $\nu \sim -1$, as seen in the inset of Fig.~\ref{vall.fig03}(b), confirming self-averaging.

In Figs.~\ref{vall.fig03}(c)-(e), we analyse self-averaging for power-law banded random matrices (PBRMs), where $\langle H_{ij}\rangle =0$ and~\cite{vall.Mirlin1996,vall.Varga2000}
\begin{equation}
\left\langle H^{2}_{ij} \right\rangle =\left \{
\begin{array}{r}
1\hspace{1.5cm}  \hspace{2cm} i=j \\
( 1+|i-j|^{2\alpha} )^{-1} \hspace{2cm} i\ne j
\end{array}
\right.
\end{equation}
PBRMs are closer to physical systems than GOE full random matrices. In PBRM, the amplitudes of the elements decrease as one moves away from the diagonal. The limits of PBRMs include: $\alpha=0$, where it coincides with full random matrices, and $\alpha \rightarrow \infty$, where the matrix converges to a tridiagonal matrix. As one increases $\alpha$ from zero, the eigenstates become less delocalized. The critical value for the transition to a localized phase is $\alpha=1$.

The results in Fig.~\ref{vall.fig03}(c), where $\alpha<1$ (delocalized regime), are similar to those for the GOE model in Fig.~\ref{vall.fig03}(b) and self-averaging is ensured. However, for $\alpha>1$ (localized regime), Fig.~\ref{vall.fig03}(d) reveals that while dephasing reduces $\overline{R_{S_{p}}}$, it fails to make it decrease as $D$ increases, so self-averaging is not achieved. In Fig.~\ref{vall.fig03}(e), we show the slope $\nu$, obtained from the scaling analysis of $\overline{R_{S_{p}}} \propto D^{\nu}$, as a function of $\alpha$. We verify that self-averaging is only observed in the delocalized regime. In the vicinity of the critical value, the fluctuations are large and $\nu>0$. Fluctuations are often larger in critical regions and our results show that even a dephasing environment cannot reduce them enough to lead to self-averaging.

In Figs.~\ref{vall.fig03}(f)-(h), we examine self-averaging in the disordered spin model given by Eq.~(\ref{vall.Eq:Ham}). The results resemble those for the PBRM. In the chaotic regime [Fig.~\ref{vall.fig03}(f)], one achieves self-averaging by opening the system, but the same is not observed when the disorder strength is large, as in Fig.~\ref{vall.fig03}(g). The analysis of the slope $\nu$ as a function of the disorder strength $h$ in Fig.~\ref{vall.fig03}(h) gives positive values for $\nu$ for any value of $h$, where level statistics as in random matrices no longer hold.

As a last remark, we reiterate that throughout this chapter, we have focused on initial states with energy near the middle of the spectrum. If instead we consider initial states close to the edges of the spectrum, even if the spin model is chaotic in the bulk, self-averaging is not achieved~\cite{vall.Vallejo2024}. This highlights the critical role of energy dependence in the study of many-body quantum dynamics, thermalization, and self-averaging, an aspect that should not be overlooked.

\section{Conclusions}

We have demonstrated that equilibration timescales depend on the observable, model, and initial state. In a chaotic disordered spin-1/2 chain with short-range interactions, the thermalization time of different observables grows polynomially with system size. This contrasts with the expectation that in chaotic many-body systems, thermalization should accelerate as the system size increases, a behaviour observed for the same observables evolving under GOE random matrices.

However, even for random matrices, observables that exhibit manifestations of spectral correlation (correlation hole) require an exponentially long time in system size to reach equilibrium. This calls for a systematic study of the equilibration time and for the eventual development of a theoretical framework. Such a study should explore the dependence of thermalization times on observables, initial state, range of the interactions, system dimension, spin size, and the role of symmetries. 

The fact that some observables, such as the survival probability, develop a correlation hole implies that experiments that probe dynamics, but have limited spectral access, can still extract information about their systems' spectra. Our proposal for detecting many-body quantum chaos through time evolution~\cite{vall.Das2025} is possible in small system sizes, where the correlation hole emerges within experimentally accessible timescales and the survival probability remains large enough for detection. These conditions make it feasible to observe signatures of quantum chaos using cold atom setups, ion traps, or commercially available quantum computers.

We also demonstrated that the lack of self-averaging in observables that exhibit the correlation hole can be circumvented by weakly coupling the system to a dephasing environment~\cite{vall.Matsoukas2023,vall.Vallejo2024}. This approach is effective in the chaotic regime, where it reduces fluctuations sufficiently to ensure self-averaging. However, near critical points, fluctuations remain too large and self-averaging is not achieved.

We used environmental coupling to suppress fluctuations and address the lack of self-averaging. However, this fluctuation reduction mechanism could have broader applications beyond self-averaging, potentially impacting a wide range of open quantum system phenomena, quantum simulation, and noise-assisted processes in quantum technologies. Investigating these applications remains an exciting avenue for future research.

\section*{Acknowledgment}

The authors thank start-up funding from the University of Connecticut.

\bibliography{biblio2025}


\renewcommand{\bibname}{References}
\begingroup
\let\cleardoublepage\relax

\endgroup


\end{document}